**Stream-K Optimization and Exploration**

**A Final Report of Stream-K Submitted to Professor Sadasivan**

**EE P 590 A: Applied Parallel Programming on GPU**

**Submitted by:**

**Casey Morrison, Nick Rackley, Bryan Gonzalez**

**5/31/2024**

https://github.com/ROCm/composable_kernel/pull/1317

## Introduction

The continuously growing demand for efficient computation in deep learning and scientific applications requires innovative approaches to optimize fundamental operations like general matrix-matrix multiplication (GEMM). This report introduces Stream-K, a novel work-centric parallel decomposition method designed to enhance GEMM performance on GPUs. Unlike traditional tile-based methods, Stream-K focuses on partitioning the workload to achieve near-perfect utilization of GPU resources, resulting in significantly improving computational efficiency and consistency.

## Background

GEMM operations are critical in various applications, ranging from deep learning to scientific simulations. To facilitate these applications scientists and engineers use GPUs that are designed to maximize throughput for such operations. However, traditional GEMM implementations face challenges in fully utilizing GPU resources due to the increasing core counts and architectural complexities of modern GPUs. The quantization inefficiency of tile-based decompositions often leads to underutilized cores and inconsistent performance, necessitating a new more flexible approach to workload distribution.

There are a couple of problems we observed with GEMM computation. Tile-based methods often face underutilization of processor cores due to mismatches in the number of tiles and cores. As you can see in Figure 1 only 75% is utilized in this example of conventional output tiles.

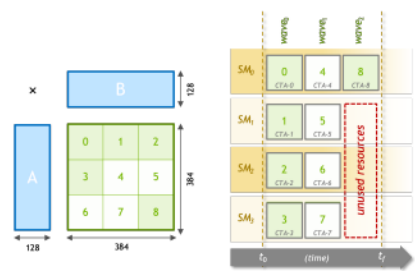

Figure 1 - Conventional tile output CU utilization (Osama et al. 2)

Traditional libraries try to mitigate this by utilizing complex kernel selection heuristics but these can struggle to maintain performance consistency due to the wide problem space. There's also an increased library size as a result of the breadth of the problem space causing file sizes to be huge, limiting portability, and creating a code maintenance problem.

Stream-K proposes solutions to all of these problems through the use of work-centric parallelization. Work is divided into smaller, even-sized work units distributed across the GPU cores as opposed to the conventional tile-based methods which would result in idle cores. Stream-K usage of even distribution of workload aims for near-perfect utilization of resources.

The algorithm design allows for one single configuration per floating point precision rather than many configurations per floating point precision. This results in improvements to storage and simplification of implementation due to there being less need for numerous pre-compiled kernel variants and also reduces the code size significantly compared to traditional methods.

## Methodology

We began by thoroughly reading the original Stream-K paper, "Stream-K: Work-centric Parallel Decomposition for Dense Matrix-Matrix Multiplication on the GPU," to understand the theoretical aspects and the algorithms proposed. Also, we examined the associated GitHub repository, focusing on the Stream-K branch, to better understand the implementation details and any deviations from the theoretical description. We then compared the details in the paper and the actual implementation looking for differences and missing pieces.

We investigated how the details in the paper held up against different hardware configurations. For instance, the paper assumes one kernel per floating point on NVIDIA hardware. We validated whether this assumption would perform similarly on other hardware such as the AMD MI100 GPU. During the investigation we measured the arithmetic intensity of 1337, indicating a large compute bottleneck in our specific application. This helped us identify the key areas where to look for computational performance optimization.

During debugging, we noticed that the Stream-K branch encountered significant errors when using the full set of commands (Ex: ./bin/example_gemm_xdl_streamk 1 2 1 30840 4096 4096 4096 4096 4096 4096 120). However, it ran correctly when excluding the final "Compute Units" parameter (Ex: ./bin/example_gemm_xdl_streamk 1 2 1). We noticed errors seemed to correlate with additional compute units being used, and realized it was likely a simple bug somewhere in the program. We resolved to find the "compute unit" bug along with other debugging and optimization efforts toward parameter optimizations. This included fine-tuning different parameters to improve performance and error minimization.

One hypothesized source of optimization would be to see the effects of removing padding, as it was not apparent from the Stream-K paper that padding was implemented, and if it is unnecessary in Stream-K, then it is quite literally artificially expanding the problem size for no benefit, in a manner particularly unattractive for GEMM, as the effects of unnecessary padding should not be uniform across all possible matrix permutations, and instead be more pronounced in permutations where padding introduces larger overhead.

## Implementation

Looking at the code base we see many implementation details that are helpful to optimizing, debugging, and working with the Stream-K algorithm.

Configuration/Testing Files
- gemm_xdl_streamk.cpp - This executable is responsible for configuring and launching the example runs of the Stream-K implementation.

- run_gemm_example.ic - This script builds the test data required for the GEMM computations and outputs the performance measurements.
- stream_config.hpp - This header file allows users to modify the number of warmup iterations and the number of runs.
- kernel_launch.hpp - This file contains the logic for launching all kernels involved in the Stream-K implementation.

Template Files:
- device_gemm_xdl_streamk.hpp - This template file includes the logic to invoke the Stream-K kernel.

Kernel File:
- gridwise_gemm_xdlops_streamk.hpp - This file contains the core logic of the Stream-K kernel.

## Results

Our investigation and implementation efforts yielded multiple insights and we saw performance improvements for the Stream-K algorithm. We thoroughly read the original Stream-K paper and subsequently the Stream-K CK library code. This helped us get a good grasp on what is going on and what is implemented vs the paper as well as finding possible areas for optimization.

We debugged the Stream-K branch to identify and resolve performance issues. Despite diving deep into the code, we couldn't track down the "compute unit bug" beyond the Block2CTile block mapping. Our contribution is still to note that running the StreamK example with default compute units (which appear to be the full MI200 120 CU's) functions fine. Realizing this, we determined it was not a critical bug, as it only affects experimentation with sub-maximal CU's, and prioritized efforts towards optimizations.

In comparing the code to the paper we found padding that was present in the code base but not in the paper. This led to an investigation into its use and its removal showed performance gains. Optimizations after many experiments with various optimizations and configurations we had most of our results here:

- Padding adjustment - Setting the padding to 0 for M, N, K dimensions resulted in 0.2%-3% improvements in performance.
    - A few suspicious results occurred during times of heavy shared use of the cluster and are disregarded from our averaged findings (IE, 3% was not a common observation, but was not obviously erroneous other than being much larger in difference between padding and no padding.
    - Curiously, a M=480, N=512, K=512 matrix failed both padded and unpadded with no other changes to the StreamK branch with 99% errors. The cause was not determined, but there is apparently a bug in the branch for this particular matrix size we did not observe with any other size.
    - Results in Table 1
- Block size adjustment - we could not get the vast majority of block/hyperparameter adjustments to compile. The CK StreamK implementation has ~15 interdependent parameters and it would take extensive learning the library or testing to even know what parameters are permissible. We did successfully compile a block size to 1024, with M and N per XDL = 16, but threw floating point errors during a run.

We then explored and spent time researching Block2Time to understand what impact this would have on the performance of Stream-K. After learning about the predictive modeling capabilities and load balancing we concluded this would lead to a performance gain in the Stream-K algorithm.

| Test | ms | Tflops | GB/s | M | N | K |
|---|---|---|---|---|---|---|
| Baseline | 1.446 | 89.07 | 66.69 | 3840 | 4096 | 4096 |
| Baseline (NP) | 1.443 | 89.26 | 66.83 | 3840 | 4096 | 4096 |
| No Padding Improvement | 0.2% | 0.2% | 0.2% | | | |
| Small matrix | 1.460 | 88.25 | 66.07 | 3 | 9 | 9 |
| Small Matrix (NP) | 1.445 | 89.12 | 66.73 | 3 | 9 | 9 |
| No Padding Improvement | 1.0% | 1.0% | 1.0% | | | |
| Ireggular Large Matrix | 0.182 | 84.10 | 127.91 | 1920 | 2000 | 2000 |
| Ireggular Large Matrix (NP) | 0.180 | 85.08 | 129.39 | 1920 | 2000 | 2000 |
| No Padding Improvement | 1.2% | 1.2% | 1.2% | | | |
| Average No Padding Improvement | 0.6% | 0.6% | 0.6% | | | |
| Medium Matrix | 99% errors | | | 480 | 512 | 512 |
| Medium Matrix (NP) | 99% errors | | | 480 | 512 | 512 |

Table 1 - Padding improvement times based on matrix size

## Conclusion

Stream – K has significant potential for optimization, making it a valuable focus for ongoing research in high-performance computing. By increasing familiarity with theoretical concepts related to GPU optimization, we can discover several performance improvements. For instance, understanding memory hierarchies, parallel processing techniques, and efficient use of resources can lead to breakthroughs in computational efficiency.

Interacting with AMD's cutting-edge software has been an insightful experience for us. It's fascinating to observe how GPUs are simplifying and accelerating complex processes. For example, in General matrix Multiply (GEMM) operations, which are fundamental to many scientific and engineering applications, the transition from CPU to GPU can reduce the operation from six nested loops to just three. This reduction is not only a testament to the efficiency of GPU architectures but also highlights the importance of optimizing algorithms to fully leverage the parallel processing power of GPUs.

Moreover, by reducing the number of computational steps and optimizing memory access patterns, we can achieve faster computation times and lower energy consumption, which is crucial when dealing with large-scale data structures.

In conclusion, the potential for optimization in Streak-K is immense. By delving into the theoretical GPU optimizations, we can drive significant advancements in computational efficiency. This not only showcases the power of modern GPUs but also paves the way for future innovations in high-performance computing.

## Future Work

Our initial pass of improving the Stream-K algorithm demonstrates promising performance improvements. However, there are several avenues for future work to enhance the efficiency and robustness of the algorithm.

The exact algorithm as put forth in the paper appears extremely compelling and seems as if it would be near optimal once fully fleshed out in implementation. Future efforts will focus on optimizing this algorithm to match the paper more stringently.

We also want to take a deeper look into different strategies to reduce the latency in hipMemcpy. By minimizing data transfer times between the host and device, we would aim to improve the efficiency of the entire Stream-K implementation.

While thinking of ways to improve development and debug time, we concluded that there needs to be more time invested in integrating automated benchmarking tools. These benchmarking tools will enable integrated and continuous performance monitoring, helping to identify bottlenecks and areas for improvement.

Finally, we would want to implement and test Block2Time which will be a major focus for future work. Utilizing Block2Time's predictive modeling capabilities, we hope to enhance the accuracy of runtime predictions and optimize the load balancing and overall performance of the Stream-K algorithm across multiple and various hardware configurations.

## Contribution of Team Members

All team members contributed equally throughout the course of the project.

Osama, Muhammad, et al. "Stream-k: Work-centric parallel decomposition for dense matrix-matrix multiplication on the gpu." *Proceedings of the 28th ACM SIGPLAN Annual Symposium on Principles and Practice of Parallel Programming*. 2023.

Sadasivan, Hari. "EE P 590A Project: Problem Statement."
https://canvas.uw.edu/courses/1708184/files/117391524?module_item_id=20297171

ROCm. "ROCm/Composable_kernel: Composable Kernel: Performance Portable Programming Model for Machine Learning Tensor Operators." *GitHub*, github.com/ROCm/composable_kernel. Accessed 31 May 2024.
7